\documentclass[12pt]{article}
\usepackage{graphicx}
\begin{document}
\small\rm
\begin{center}
{\Large\bf DANSSino: a pilot version of the DANSS neutrino detector}\\[2mm]
{\em
{I.~Alekseev~$^a$},
{V.~Belov~$^b$},
{V.~Brudanin~$^b$},
{M.~Danilov~$^{a,c}$},
{V.~Egorov~$^b$\footnote{Corresponding author; e-mail:  egorov@jinr.ru}},
{D.~Filosofov~$^b$},
{M.~Fomina~$^b$},
{Z.~Hons~$^b$},
{A.~Kobyakin~$^a$},
{D.~Medvedev~$^b$},
{R.~Mizuk$^a$},
{E.~Novikov~$^a$},
{A.~Olshevsky~$^b$},
{S.~Rozov~$^b$},
{N.~Rumyantseva~$^b$},
{V.~Rusinov~$^a$},
{A.~Salamatin~$^b$},
{Ye.~Shevchik~$^b$},
{M.~Shirchenko~$^b$},
{Yu.~Shitov~$^{b,d}$},
{A.~Starostin~$^a$},
{D.~Svirida~$^a$},
{E.~Tarkovsky~$^a$},
{I.~Tikhomirov~$^a$},
{E.~Yakushev~$^b$},
{I.~Zhitnikov~$^b$},
{D.~Zinatulina~$^b$}
}\\[2mm]
$^a$ {\scriptsize\rm ITEP -- State Scientific Center, Institute for
 Theoretical and Experimental Physics, Moscow, Russia}\\
$^b$ {\scriptsize\rm JINR -- Joint Institute for Nuclear Research, Moscow Region, Dubna, Russia}\\
$^c$ {\scriptsize\rm MIPT -- Moscow Institute of Physics and Technology, Moscow Region, Dolgoprudny, Russia}\\
$^d$ {\scriptsize\rm ICL -- Imperial College London, SW7 2AZ, London, United Kingdom}
\end{center}
\begin{flushleft}
\end{flushleft}
\noindent
{\scriptsize DANSSino is a reduced pilot version of a solid-state detector of reactor antineutrinos (to be created within the DANSS project and installed under the industrial 3~GW$_{\rm th}$ reactor of the Kalinin Nuclear Power Plant -- KNPP). Numerous tests performed at a distance of 11 m from the reactor core demonstrate operability of the chosen design and reveal the main sources of the background. In spite of its small size ($20\times20\times100$~cm$^3$), the pilot detector turned out to be quite sensitive to reactor antineutrinos, detecting about 70 IBD events per day with the signal-to-background ratio about unity.}

\newpage
\section{Detector DANSS}
\begin{figure}[ht]
 \input{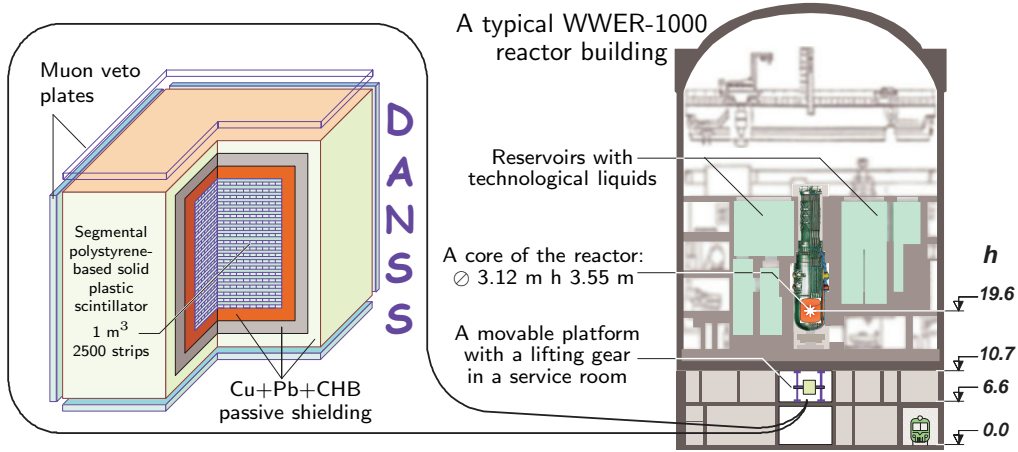}
 \caption{\footnotesize Design of the DANSS neutrino detector (left) and its position under the industrial reactor WWER-1000 (right).}
 \label{Fig.DANSS}
\end{figure}

The DANSS project \cite{DANSS_TAUP,DANSS_Viena,DANSS_ICHEP} is aimed at creating a relatively compact neutrino spectrometer which does not contain any flammable or other dangerous liquids and may therefore be located very close to the core of an industrial power reactor (Fig.~\ref{Fig.DANSS}). Due to a high neutrino flux ($\sim5\times10^{13}\; \bar\nu_e /{\rm cm}^2/{\rm s}$ at a distance of 11 m) it could be used for the reactor monitoring and for fundamental research including neutrino oscillation studies.

The DANSS detector will consist of highly segmented plastic scintillator with a total volume of 1 m$^3$, surrounded with a composite shield of copper (Cu), lead (Pb) and borated polyethylene (CHB), and vetoed against cosmic muons with a number of external scintillator plates.

\begin{figure}[th]
 \input{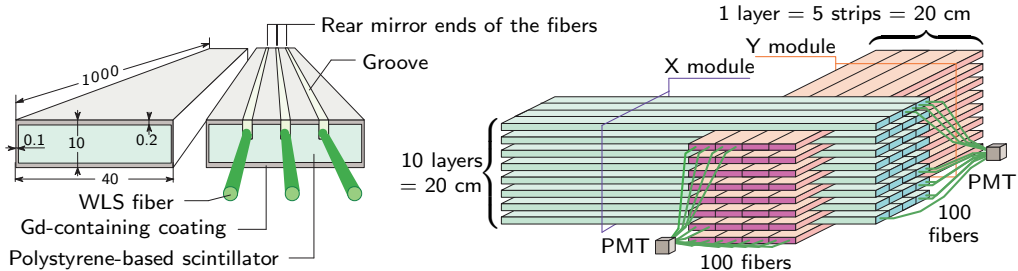}
 \caption{\footnotesize The basic element (left) and two of fifty intersecting modules (right) of the DANSS detector.}
 \label{Fig.DANSS_Modules}
\end{figure}

The basic element of DANSS is a polystyrene-based extruded scintillator strip ($1\times4\times100$~ cm$^3$) with a thin Gd-containing surface coating which is a light reflector and an ($n,\gamma$)-converter simultaneously (Fig.~\ref{Fig.DANSS_Modules}).
The coating (about 0.1--0.2~mm) is produced by co-extrusion and consists of polystyrene with 18\% admixture of rutile and 6\% of gadolinium oxide, so that the final Gd density is about 1.6~mg/cm$^2$, which corresponds to $\sim$0.35\%$_{\rm wt}$. Light collection from the strip is done via three wavelength-shifting Kuraray fibers Y-11, $\oslash$~1.2~mm, glued into grooves along all the strip. An opposite (blind) end of each fiber is polished and covered with a mirror paint, which decreases a total lengthwise attenuation of a light signal down to $\sim$5~\%/m.

Each 50 parallel strips are combined into a module, so that the whole detector (2500 strips) is a structure of 50 intercrossing modules (Fig.~\ref{Fig.DANSS_Modules}). Each module is viewed by a compact photomultiplier tube (Hamamatsu R7600U) coupled to all 50 strips of the module via 100 WLS fibers, two per strip. In addition, to get more precise energy and space pattern of an event, each strip is equipped with an individual multipixel photosensor (SiPM) operating in the Geiger mode and coupled to the strip via the third WLS fiber.

In order to check operability of the DANSS design, compare different acquisition schemes and measure the real background conditions, a simplified pilot version of the detector (DANSSino) was created. Below we present the DANSSino description and some results of numerous tests performed at the JINR laboratory, as well as under the KNPP reactor.

\section{The DANSSino design}
\begin{figure}[ht]\label{Fig.DANSSino}
 \input{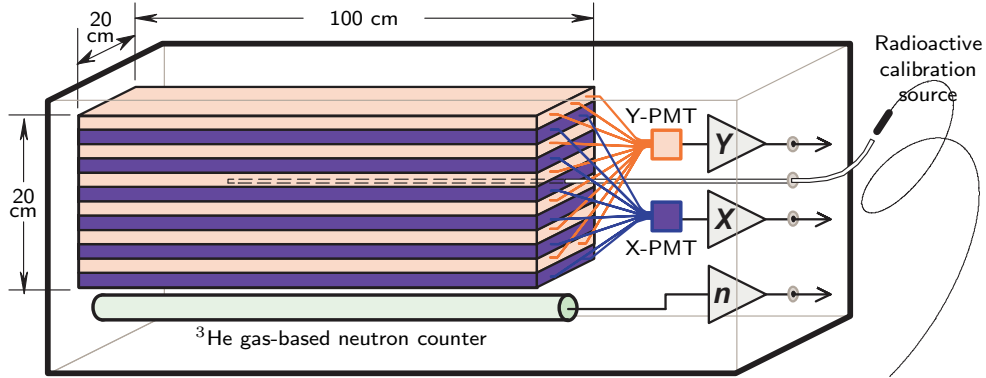}
 \caption{\footnotesize The DANSSino detector}
 \end{figure}

DANSSino consists of exactly the same basic elements as the main DANSS detector.
One hundred strips of DANSSino (Fig.~\ref{Fig.DANSSino}) form a bar $20\times20\times100$ cm$^3$ divided into two modules: the odd strip layers are coupled to the X-PMT and the even ones to the Y-PMT. Together with an additional neutron counter\footnote{This $^3$He gas-based counter allows high sensitivity monitoring of the thermal neutron flux inside the detector shielding.} both modules are equipped with simple front-end preamplifiers and placed into a light-tight box. Information from the individual photodiodes of each strip is not used in this prototype.
To perform energy calibration, a teflon tube is placed inside the bundle of strips, so that a tiny radioactive source can be inserted in the detector with a thin flexible string. For this purpose several gamma and neutron sources with activity of few Bq were produced: $^{137}$Cs, $^{60}$Co, $^{22}$Na, $^{248}$Cm. To suppress external background caused by gamma-rays and thermal neutrons the detector is surrounded with a passive shield. As it is rather compact, the composition of the shielding can be easily changed. A set of big scintillator plates ($200\times50\times3$~cm$^3$) similar to that employed in the GERDA experiment \cite{GERDA} form an active veto system which is used to tag the events associated with cosmic muons.

The Inverse Beta-Decay (IBD) of hydrogen atoms in the detector body is used to detect the reactor antineutrino
\begin{equation}
 \widetilde{\nu_e}+p\rightarrow e^++n\;.
\end{equation}
The detection process proceeds in two steps: the first one applies to the positron and the second to the neutron. The energy threshold of the IBD reaction is 1.8~MeV, while most of the remaining neutrino energy is transferred to the positron. The positron deposits its energy within a short range of few cm and then annihilates emitting two 511~keV photons at 180$^\circ$. As a result, the first (Prompt) energy deposit is distributed in space in a very specific way. The second (Delayed) step is the detection of the neutron. Initial energy of the neutron is only few keV. After moderation in the plastic scintillator it is captured by $^{157}$Gd or $^{155}$Gd with a very high cross-section. In both cases a cascade of $\gamma$-rays is emitted with the total energy of about 8~MeV. Because of high multiplicity and deep penetration in plastic these $\gamma$-rays produce a flash which is spread widely within a sphere with a diameter of about 30-40~cm, so that a number of strips in both X and Y modules are usually fired. Distribution of time between the Prompt and Delayed signals is described by a combination of two exponents
 \begin{equation} \label{Eq.f(t)}
 f_1(t)=\frac1{\tau_c-\tau_m}\left(e^{-t/\tau_c}-e^{-t/\tau_m}\right)\;,
 \end{equation}
where the characteristic times $\tau_m$ and $\tau_c$ correspond to the neutron moderation and capture respectively and depend on the detector structure.

Though the IBD event has a very specific signature, it occurs under intense external and internal $\gamma$, $n$ and $\mu$ background. Therefore, adequate selection rules ({\sl hardware trigger}) should be worked out in advance.

 \begin{figure}[ht]
 \input{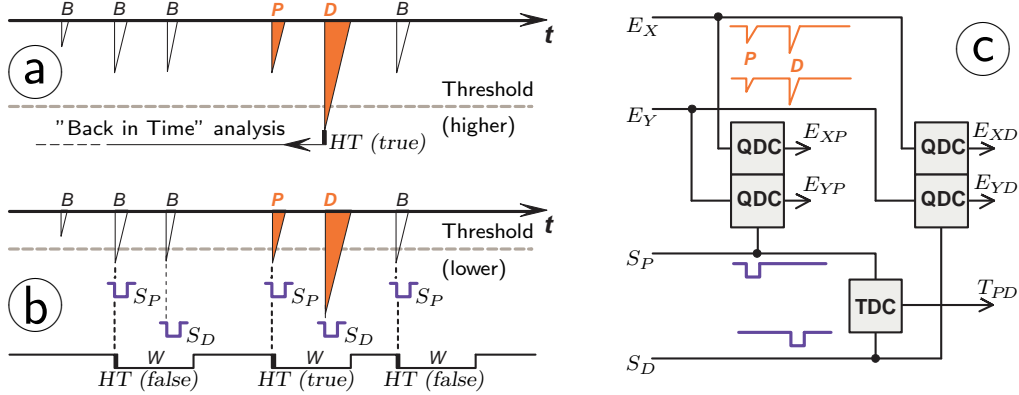}
 \caption{\footnotesize Left: two alternative types of the hardware trigger~($HT$) for the true IBD event consisting of the Prompt ($P$) and Delayed ($D$) signals in presence of background pulses ($B$). Right: a simplified diagram of the QDC-based acquisition system used in DANSSino.}\label{Fig.ACQ}
 \end{figure}

There are two obvious types of this hardware trigger. The most reliable of them is detection of the Delayed neutron capture, as the amplitude and multiplicity of the neutron signal are much higher than those of the natural $\gamma$-background. This method  (Fig.~\ref{Fig.ACQ}a) requires  digitization of the total data stream with flash ADCs and subsequent on-line analysis of the preceding signals (in this way one can spot the Prompt signal which could happen few tens of a microsecond before the hardware trigger $HT$ and have relatively low energy). This trigger is planned to be used in the full-scale DANSS spectrometer, but at the DANSSino stage it seems to be untimely and impractical.

The other trigger implemented in this work is simpler but less reliable because it can successfully function only under a lower background count rate. Within this method (Fig.~\ref{Fig.ACQ}b), the hardware trigger $HT$ is produced by any Prompt signal, and then the system waits for the Delayed signal during some fixed time $W$. The energy of both Prompt and Delayed signals ($E_P$ and $E_D$) detected by both X and Y modules are measured with two Charge-to-Digital Converters (QDC$_P$ and QDC$_D$), which are gated separately by the $S_P$ and $S_D$ strobes (Fig.~\ref{Fig.ACQ}c). Finally, each collected event contains 4 energies ($E_{XP}$, $E_{YP}$, $E_{XD}$, $E_{YD}$), time between the P and D pulses ($T_{PD}$) and information about the muon veto (which of the plates were fired and when).

\section{Ground laboratory tests}
\begin{figure}[ht]
 \input{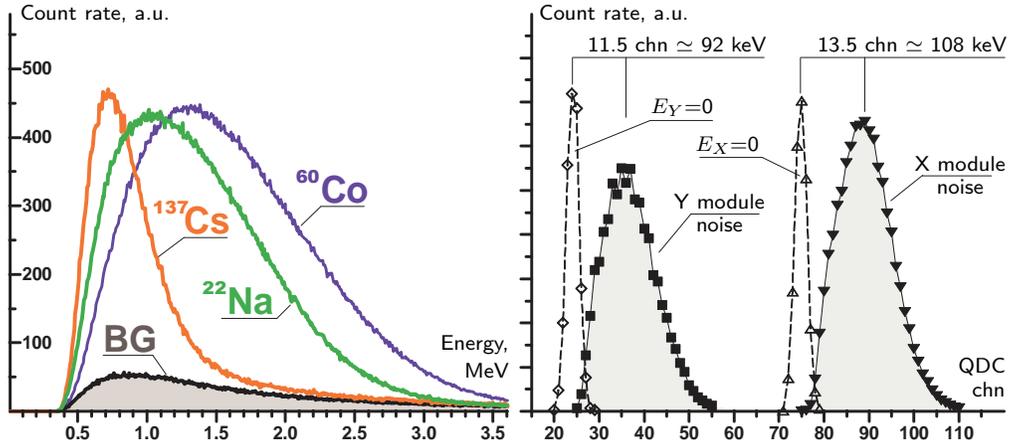}
 \caption{\footnotesize Energy spectra measured with $\gamma$-sources (left) and the noise peaks (right).}\label{Fig.gamma}
 \end{figure}

Numerous tests with the DANSSino detector were performed at the ground floor of the JINR laboratory building (Dubna). Measurements with $^{137}$Cs, $^{22}$Na and $^{60}$Co $\gamma$-sources (Fig.~\ref{Fig.gamma}) compared with the MC simulations were used to calibrate the energy scale, estimate the energy resolution and set the lower energy thresholds of the discriminators.

As the light collection from a long strip is far from being perfect, the number of photoelectrons (p.e.) produced at the PMT photocathode is relatively small. Therefore, the energy resolution of the spectrometer is determined mainly by the Poisson dispersion of this number and depends on the energy deposit in the scintillator which corresponds to a single photoelectron $\varepsilon_{\rm pe}$. In order to estimate it, single-photoelectron noise peaks were measured with both X and Y modules. As a result, $\varepsilon_{\rm pe}$ was found to be about 100 keV/p.e.

\begin{figure}[ht]
 \input{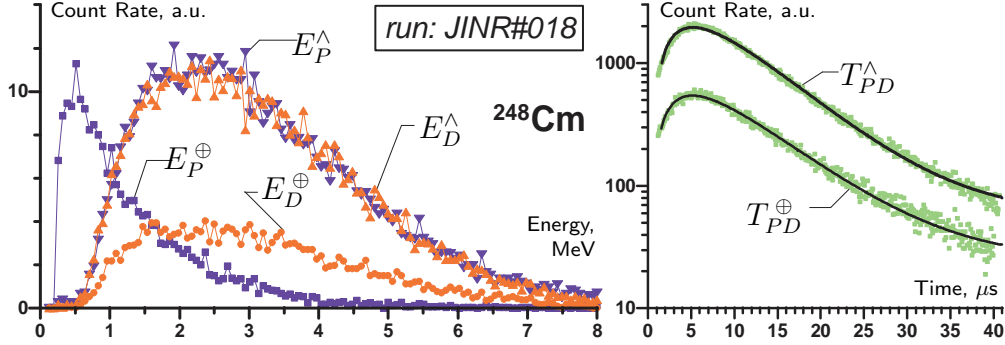}
 \caption{\footnotesize Energy (left) ant time (right) spectra of the neutrino-like events measured with the $^{248}$Cm source. The spectra corresponding to the ($X_P\oplus Y_P$) and ($X_P\wedge Y_P$) logic of the Prompt signals are marked with $\oplus$ and $\wedge$ superscripts, respectively. The Delayed signal is required to be of the ($X_D\wedge Y_D$) type only. }\label{Fig.248Cm}
 \end{figure}

The next important parameter of the spectrometer is its selective sensitivity to neutrons. The $^{248}$Cm neutron source provides signals with a signature very similar to the neutrino-like events. Indeed, the Delayed signal is caused by the same $^{157,155}$Gd($n,\gamma$) reaction, whereas the Prompt signal corresponds either to the ($n$-$p$) scattering of the initial fast neutron on hydrogen in the plastic detector or to the prompt $\gamma$-rays following the $^{248}$Cm fission. In the first case the most probable is the response of only one of the two PMTs ($X_P\oplus Y_P$) and in the second case the response of both ($X_P\wedge Y_P$)\footnote{The symbols $\wedge$ ({\sf AND}) and $\oplus$ ({\sf XOR}) here and below stand for logical operations of conjunction and exclusive disjunction, respectively.}. The Prompt energy spectra corresponding to these cases are very different, whereas the Delayed spectra have the same shape (Fig.~\ref{Fig.248Cm}).

The $T_{PD}$ time distribution is also very similar to the IBD. The only difference is much higher neutron multiplicity ($k\simeq3$), so that instead of expression (\ref{Eq.f(t)}) it is described by a more complicated function
\begin{equation} \label{Eq.fk(t)}
f_k(t)=f_1(t)\cdot\!\left(1\!-\!\!\int\limits_0^t \!f_1(\tau)d\tau \right)^{\!k-1}\!\!\!\!\!\!=\;\;
\frac{e^{\frac{-t}{\tau_c}}-e^{\frac{-t}{\tau_m}}}{\left(\tau_c-\tau_m\right)^k}\cdot
\left(\tau_c\, e^{\frac{-t}{\tau_c}}-\tau_m\, e^{\frac{-t}{\tau_m}}\right)^{k-1},
\end{equation}
which reflects the fact than only the first of $k$ neutrons is detected by our acquisition system.  Quantitative analysis of the $^{248}$Cm data shown in Fig.~\ref{Fig.248Cm} with $k=3$ gives the following values for the moderation and capture specific times:
\begin{equation} \label{Eq.tau(Cm)}
\tau_m\simeq3\pm1\;{\rm \mu s}\hspace{20mm}
\tau_c\simeq24\pm1\;{\rm \mu s}
\end{equation}

Measurements performed without radioactive sources but with different composition of passive shielding confirm that 10~cm of heavy material (copper and/or lead) is quite enough to suppress the natural gamma background, and 10~cm of borated polyethylene reject thermal neutrons. On the other hand, it was found that in addition to gamma rays, muons and thermal neutrons, the laboratory natural background contains a significant number of neutrino-like events ($\simeq300$ per hour) consisting of both Prompt and Delayed signals -- (P+D)-pairs. The $E_D$ energy spectrum and $T_{PD}$ time distribution of those events are of the similar shape as in the $^{248}$Cm case. It is natural to assume that the events are caused by fast neutrons which could not be rejected even with 16~cm of borated polyethylene. Indeed, it is known \cite{Heusser,AtmosphereBG} that the high-energy hadronic component of the cosmic background at sea level could be as high as $10^2$/m$^2$/s depending on the energy and the site\footnote{More precise calculation \cite{FluxCalculator} for neutrons with $E_n\geq10$~MeV under the JINR conditions yield the value of (40-50)/m$^2$/s.}, and a low-Z shielding equivalent to at least 20 meters of water (20 m w. e.) is needed to suppress it.

\section{Tests under the KNPP industrial reactor}
After the ground laboratory tests the DANSSino detector was transported to the Kalinin Nuclear Power Plant (Udomlya, 285 km NW of Dubna) and installed in a service room of Unit\#3, at a distance of 11 m from the reactor core center. Unit\#3 is a twin of Unit\#4 where the main DANSS detector is being mounted now (Fig.~\ref{Fig.DANSS}).
The unit includes a typical Russian industrial water-moderated water-cooled power reactor WWER-1000 \cite{WWER-1000} of thermal power 3~GW$_{th}$. Huge reservoirs with technological liquids, thick walls of heavy concrete, the reactor body and equipment placed above the room provide excellent shielding ($\simeq$50 m w. e.) which completely removes fast cosmic neutrons. The muon component is suppressed by a factor of $\simeq$6. The gamma background at the detector site is slightly higher than at the JINR laboratory because of high $^{40}$K contamination of the surrounding concrete. Thermal and epithermal reactor neutrons penetrate the room through the monitoring tubes and increase the average background as well. Fortunately, the flux of those neutrons measured with the external four-fold $^3$He gas-based neutron detector turned out to be not too high and could be suppressed with $\sim$10--15~cm of borated polyethylene (CHB); in addition, slow neutrons cannot emulate the IBD process because of their small energy.

The results of some background measurements performed at the JINR laboratory and under the operating KNPP reactor with and without shielding are presented in Table~\ref{Tab.BG}. Some of the JINR and KNPP measurement conditions, such as the PMT high voltage, discriminator thresholds, energy scale, etc., could be slightly different, so that the values presented in the table should be considered as illustrative only. The X and Y columns of the table indicate the number of raw pulses with an energy $\geq$ 0.25 MeV. From these values it is seen that although the initial background conditions under the reactor are worse, a relatively thin passive shielding (10 cm of lead and 16 cm of borated polyethylene) improves it significantly and makes even 3 times better than in the laboratory. The next two columns reflect mainly the flux of thermal neutrons and cosmic muons, respectively (the neutrons produce X$\wedge$Y-coincidences with the total energy between the threshold and 8 MeV, whereas the muons also cause X$\wedge$Y coincidences but saturate the QDCs).

\begin{table}[ht]
\caption{\footnotesize Background in the JINR laboratory and under the KNPP reactor measured by DANSSino unshielded and shielded with 10 cm of lead, 16 cm of CHB and $\mu$-veto plates (here and below the shielding composition is enumerated from inside to outside).}
\label{Tab.BG}
{\footnotesize
\begin{tabular}{|c|c||r|r||r|r||r|r|} \hline
Operation  & Detector \rule{0mm}{4mm}&\multicolumn{4}{|c||}{Module count rate,}
&\multicolumn{2}{|c|} {(P+D) pairs}\\
conditions & shielding \rule[-2mm]{0mm}{3mm}&\multicolumn{4}{|c||}{counts per second}
&\multicolumn{2}{|c|} {per day}\\ \cline{3-8}
&&\multicolumn{1}{|c|}{\scriptsize\sf X}
&\multicolumn{1}{|c||}{\scriptsize\sf Y}
&\multicolumn{1}{|c|}{\scriptsize\sf X$\wedge$Y}
&\multicolumn{1}{|c||}{\scriptsize\sf X$\wedge$Y}&&\\
&&\multicolumn{1}{|c|}{\scriptsize\sf E$\geq$0.25}
&\multicolumn{1}{|c||}{\scriptsize\sf E$\geq$0.25}
&\multicolumn{1}{|c|}{\scriptsize\sf E$\geq$0.5}
&\multicolumn{1}{|c||}{\scriptsize\sf E$\geq$8.0}
&\multicolumn{1}{|c|}{{\scriptsize\sf NO }$\mu$}
&\multicolumn{1}{|c|}{{\scriptsize\sf AND }$\mu$}\\
&&\multicolumn{1}{|c|}{\scriptsize\sf$\gamma$+$n$+$\mu$}
&\multicolumn{1}{|c||}{\scriptsize\sf$\gamma$+$n$+$\mu$}
&\multicolumn{1}{|c|}{\scriptsize\sf$n$+$\mu$}
&\multicolumn{1}{|c||}{\scriptsize\sf$\sim\mu$}&&\\ \hline\hline
  JINR& no shielding  \rule[-1.5mm]{0mm}{5.0mm}& 532 & 465 & 235  & 19 & \multicolumn{2}{|r|}{601 400\hspace*{5mm}} \\ \cline{2-8}
{\sf natural BG}\rule[-1.5mm]{0mm}{5.0mm}      & {\scriptsize\sf Pb+CHB+$\mu$-veto} & 61 & 58 & 42 & 17 & 30 750 & 9 030 \\ \hline \hline
 KNPP & no shielding \rule[-1.5mm]{0mm}{5.0mm} & 1 470 & 1 360 & 408  & 4 & \multicolumn{2}{|r|}{11 837 500\hspace*{5mm}} \\ \cline{2-8}
 {\scriptsize\sf $5\!\times\!10^{13}\nu$/cm$^2\!$/s}\rule[-1.5mm]{0mm}{5.0mm}&{\scriptsize\sf Pb+CHB+$\mu$-veto}& 20 & 19 & 11 & 2 & 1 240 & 980 \\ \hline
\end{tabular}
}

\end{table}

Finally, the last two columns of Table~\ref{Tab.BG} show the rate of events consisting of (Prompt+Delayed) signal pairs without any additional selection. A big number of these events for the unshielded detector arises from random coincidences caused by the high raw count rate. In the case of the shielded detector those are signals mainly from the fast neutrons. As expected, the number of these false neutrino-like events under the reactor becomes orders of magnitude lower than at the laboratory, but still differs from zero.
The remaining part is mostly associated with the $\mu$-veto signals and corresponds probably to secondary fast neutrons produced by cosmic muons in the surrounding heavy materials (lead and copper).

\begin{table}[hb]
\caption{\footnotesize Number of false neutrino-like events associated with $\mu$-veto signals and registered by DANSSino shielded with different materials.} \label{Tab.NuMu(shld)}
\begin{center}
{\footnotesize
\begin{tabular}{|l||c|c|c|} \hline
{\sf Run\#}
& \multicolumn{1}{|c|}{\sf KNPP\#027}
& \multicolumn{1}{|c|}{\sf KNPP\#035}
& \multicolumn{1}{|c|}{\sf KNPP\#044}\\
{\sf Shielding}
& \multicolumn{1}{|c|}{\scriptsize\sf \rule[-5mm]{0mm}{10mm}
  \parbox{18mm}{Pb \hfill{ }=10 cm\\CHB \hfill{ }=16 cm}}
& \multicolumn{1}{|c|}{\scriptsize\sf
  \parbox{18mm}{CHB \hfill{ }=8 cm\\Pb \hfill{ }=10 cm}}
& \multicolumn{1}{|c|}{\scriptsize\sf
  \parbox{18mm}{Cu \hfill{ }=5 cm\\CHB \hfill{ }=8 cm\\Pb \hfill{ }=5 cm}}
\\ \hline
{\sf Events per day}\rule[-1.5mm]{0mm}{5.0mm}& 835$\pm$7 & 214$\pm$4 & 384$\pm$4\\ \hline
\end{tabular}
}
\end{center}

\end{table}

This assumption was confirmed by measurements performed with different composition of the passive shielding (Table~\ref{Tab.NuMu(shld)}). These data were analyzed under relatively strong requirements which correspond to the IBD signature:
\begin{itemize}
\item the time between the Prompt and Delayed signals must be within a reasonable range $T_{PD}\in \left[1.5-30.0\right]\,\mu$s;
\item the Delayed signal should correspond to the Gd($n$,$\gamma$) reaction, i.e., both the X and Y modules should be fired ($X_D\wedge Y_D$) with a reasonable\footnote{As the detector is small, significant part of the $\gamma$-cascade is not detected, and therefore the acceptable $E_D$ range is extended to the lower energy.} total energy $E_{XD}+E_{YD}=E_D \in\left[1-8\right]$~MeV;
\item the Prompt energy must also be in a right range $E_P \in\left[1-7\right]$~MeV.
\end{itemize}

It is evident from Table~\ref{Tab.NuMu(shld)} that 5 cm of copper in the close vicinity to the detector body increase the number of false neutrino-like events by a factor of 2, and 10 cm of lead increase it by a factor of 4. Of course, these false events could be tagged and then rejected by the veto system, but the efficiency of any veto never reaches 100\%. On the other hand, heavy materials cannot be avoided completely as they are needed for shielding against external gamma-rays; therefore they should be isolated from the detector with an intermediate moderator.

\begin{figure}[hb]
 \input{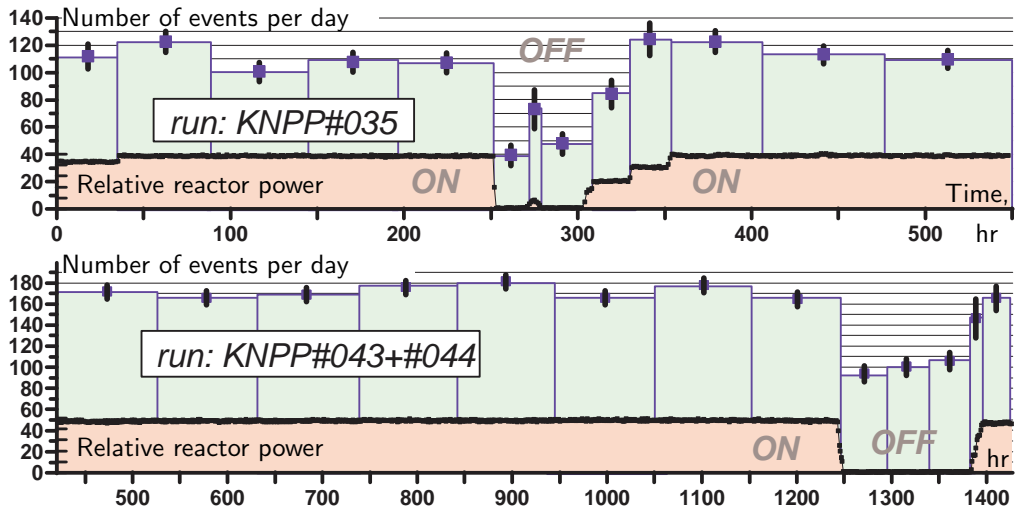}
 \caption{\footnotesize Time dependence of the reactor power (bottom of each diagram) and the number of the neutrino-like events detected by DANSSino (top of each diagram) for two measurement periods.}\label{Fig.Nu(time)}
\end{figure}

Two of our measurement runs came across few-days interruptions in the reactor operation (OFF periods), so that it was possible to estimate background experimentally without the neutrino flux\footnote{Really, when the reactor is stopped, neutrinos are still emitted by the radioactive fission products, but the decay energy of long-lived nuclei is low and therefore those OFF-neutrinos cannot be detected because of the IBD energy threshold.}. Figure~\ref{Fig.Nu(time)} shows time diagrams of these runs. The lower parts of the diagrams show the actual reactor power measured with the external high-sensitive neutron counter installed near the outlet of the monitoring neutron tube. Strong correlation between the reactor power and the number of the neutrino-like events detected by DANSSino is obvious.

 \begin{figure}[ht]
 \input{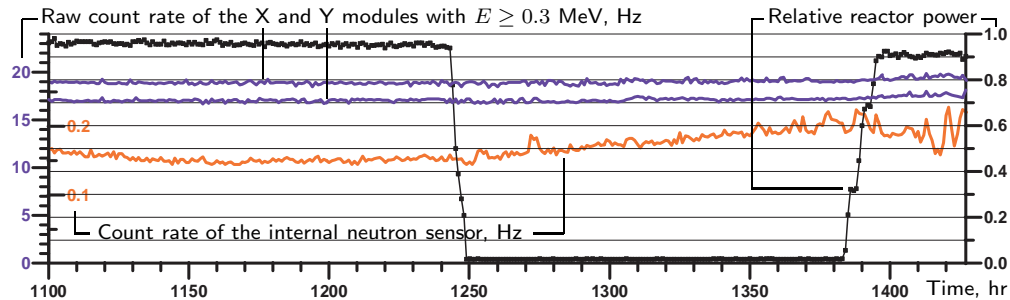}
 \caption{\footnotesize Time evolution of the reactor power during Run KNPP\#044 and the raw count rate of the neutron sensor inside the shielding and both the X and Y modules. No visible correlation is observed.}\label{Fig.CR(time)}
 \end{figure}

There could be doubts about a nature of the count rate variations -- are they really caused by the neutrino-induced IBD or by the random coincidences of gamma-rays with external neutron background, which is known to be strongly correlated with the reactor operation? Trying to answer this question, we analyzed time evolution of the background inside the DANSSino shielding (Fig.~\ref{Fig.CR(time)}). At first glance, it seems that there is no visible correlation between the reactor power and the raw count rate of both X and Y modules (which detect mainly gamma rays). Unfortunately, sensitivity of the $^3$He gas-based neutron sensor is not high enough to detect a very low but still non-zero neutron flux inside the shielding. Therefore, its count rate shown in Fig.~\ref{Fig.CR(time)} is caused mainly by cosmic rays and internal radioactive pollution, as well as by electric noise, so that absence of any visible correlation cannot be considered as an argument in favour of the neutrino versus the neutron correlated background.

Numerical values of the neutrino-like count rates, Signal-to-Background ratios and muon-induced background rates are given in Table~\ref{Tab.Nu(ON-OFF)}, which describes the neutrino-like events detected with different low energy thresholds ($E_P^{\rm min}$=0.5, 1.0 and 1.5 MeV) and alternative logic of the Prompt signals ($X_P\oplus Y_P$) and ($X_P\wedge Y_P$).

\begin{table}[th]
\caption{\footnotesize Rate of the neutrino-like events with ($X_P\oplus Y_P$) and ($X_P\wedge Y_P$) logic of the Prompt signal registered by DANSSino at the ON and OFF reactor periods.}
\label{Tab.Nu(ON-OFF)}
{\small\rm
\begin{tabular}{|c|c|c||r|r||r|r|r|r|} \hline
{\sf RUN\#}&\multicolumn{2}{|c||}{$P$-signal}&
\multicolumn{5}{|c|}{\sf Number of neutrino-like events per day}&
\multicolumn{1}{|c|}{\sf  S/B } \\ \cline{2-8}
 {\scriptsize\sf shielding}
 &{\scriptsize $E_P^{\rm min\!}$}&{\scriptsize\sf $X_P$-$Y_P$} &
\multicolumn{2}{|c||}{\footnotesize\sf tagged by $\mu$-veto} &
\multicolumn{2}{|c|} {\footnotesize\sf free of $\mu$-veto} &
\multicolumn{1}{|c|}{\sf S =} &
\multicolumn{1}{|c|}{\sf ratio}\\ \cline{4-7}
 {\scriptsize\sf structure}&
 {\scriptsize\sf MeV}&{\footnotesize\sf logic}&
\multicolumn{1}{|c|}{\footnotesize\sf ON} &
\multicolumn{1}{|c||}{\footnotesize\sf OFF} &
\multicolumn{1}{|c|}{\footnotesize\sf ON} &
\multicolumn{1}{|c|}{\footnotesize\sf OFF} &
\multicolumn{1}{|c|}{\footnotesize\sf ON--OFF} &
\multicolumn{1}{|c|}{\footnotesize ($\approx$)} \\ \hline \hline
{\scriptsize\sf \hspace*{-2mm}KNPP\#035\hspace*{-2mm}}
 &0.5&{\scriptsize\bf $X_p\!\oplus\! Y_p$}&
       169$\pm$3 & 173$\pm$9 &361$\pm$4 &249$\pm$11$\!$&112$\pm$12$\!$ &0.45\\ \cline{3-9}
 &   &{\scriptsize\bf $X_p\!\wedge\! Y_p$}&
       129$\pm$3 & 133$\pm$8 & 69$\pm$2 & 38$\pm$~4 & 30$\pm$~5 &0.79\\ \cline{2-9}
 {\scriptsize\sf CHB \hfill{ }8 cm}
 &1.0&{\scriptsize\bf $X_p\!\oplus\! Y_p$}&
        92$\pm$2 &  95$\pm$7 &125$\pm$3 & 66$\pm$~6 & 59$\pm$~6 &0.89\\ \cline{3-9}
 {\scriptsize\sf Pb  \hfill{ }10 cm}
 &   &{\scriptsize\bf $X_p\!\wedge\! Y_p$}&
       123$\pm$3 & 126$\pm$8 & 60$\pm$2 & 31$\pm$~4 & 29$\pm$~4 &0.92 \\ \cline{2-9}
 &1.5&{\scriptsize\bf $X_p\!\oplus\! Y_p$}&
        59$\pm$2 &  63$\pm$6 & 62$\pm$2 & 29$\pm$~4 & 34$\pm$~4 &1.16\\ \cline{3-9}
 &   &{\scriptsize\bf $X_p\!\wedge\! Y_p$}&
       108$\pm$2 & 105$\pm$7 & 48$\pm$2 & 20$\pm$~3 & 28$\pm$~4 &1.43\\ \hline\hline
{\scriptsize\sf \hspace*{-2mm}KNPP\#043\hspace*{-2mm}}
 &0.5&{\scriptsize\bf $X_p\!\oplus\! Y_p$}&
       301$\pm$4 & 300$\pm$7 &487$\pm$5 &401$\pm$~9 & 86$\pm$10$\!$ &0.44\\ \cline{3-9}
{\scriptsize\sf + \#044} &   &{\scriptsize\bf $X_p\!\wedge\! Y_p$}&
       243$\pm$3 & 246$\pm$7 &112$\pm$2 & 76$\pm$~4 & 36$\pm$~4 &0.46\\ \cline{2-9}
 {\scriptsize\sf Cu \hfill{ }5 cm}
 &1.0&{\scriptsize\bf $X_p\!\oplus\! Y_p$}&
       156$\pm$3 & 158$\pm$5 &188$\pm$3 &130$\pm$~5 & 58$\pm$~6 &0.44\\ \cline{3-9}
 {\scriptsize\sf CHB\hfill{ }8 cm}
 &   &{\scriptsize\bf $X_p\!\wedge\! Y_p$}&
       228$\pm$3 & 231$\pm$6 & 99$\pm$2 & 64$\pm$~3 & 35$\pm$~4 &0.54\\ \cline{2-9}
 {\scriptsize\sf Pb \hfill{ }5 cm}
 &1.5&{\scriptsize\bf $X_p\!\oplus\! Y_p$}&
        94$\pm$2 &  99$\pm$4 & 93$\pm$2 & 52$\pm$~3 & 42$\pm$~4 &0.80\\ \cline{3-9}
 &   &{\scriptsize\bf $X_p\!\wedge\! Y_p$}&
       195$\pm$3 & 201$\pm$6 & 77$\pm$2 & 44$\pm$~3 & 33$\pm$~3 &0.74\\ \hline
\end{tabular}
}

\end{table}

Analysis of Table~\ref{Tab.Nu(ON-OFF)} allows the following conclusions.
\begin{itemize}
\item
Intensities of the muon-induced background events do not differ for the ON and OFF periods, which confirms the adequate operation of the acquisition system.
\item
A heavy shielding (5 cm of copper) without an inner moderator increases the background by 70\%, as was already shown in Table~\ref{Tab.NuMu(shld)}, but improves the efficiency\footnote{The efficiency here means the probability to detect the IBD if it occurs.} by $\sim$10\%, returning part of the escaped true IBD neutrons to the scintillator.
\item
The ($X_P\wedge Y_P$) logic of the Prompt signal corresponds to the stronger selection condition and provides better S/B ratios but poorer statistics. Indeed, the range of the recoil proton after the neutron scattering in poly\-sty\-rene does not exceed 1--2~mm and therefore the Prompt signal of the false background event caused by a fast neutron is detected most probably by {\sl only one} of the two modules ($X_P\!\oplus\! Y_P$).
\item
Lowering the Prompt low-energy threshold $E_P^{\rm min}$ below $\sim$1~MeV in case of ($X_P\!\wedge\! Y_P$) logic gives no visible gain in statistics and only increases the background thus impairing the S/B ratio.
\item
Under less strong ($X_P\!\oplus\! Y_P$) condition this lowering also gives rise to the probability of random coincidences of $\gamma$-rays with external thermal neutrons. It so happened that the detector in the considered test runs was not shielded against neutrons well enough, and therefore the rate of such random coincidences was higher in the ON period, thus generating a false positive (ON--OFF) difference. As the external $\gamma$-rays produce mostly low-amplitude single pulses, the above effect shows itself mainly at the low $E_P$ under the ($X_P\!\oplus\! Y_P$) condition.
\end{itemize}

\begin{figure}[ht]
 \input{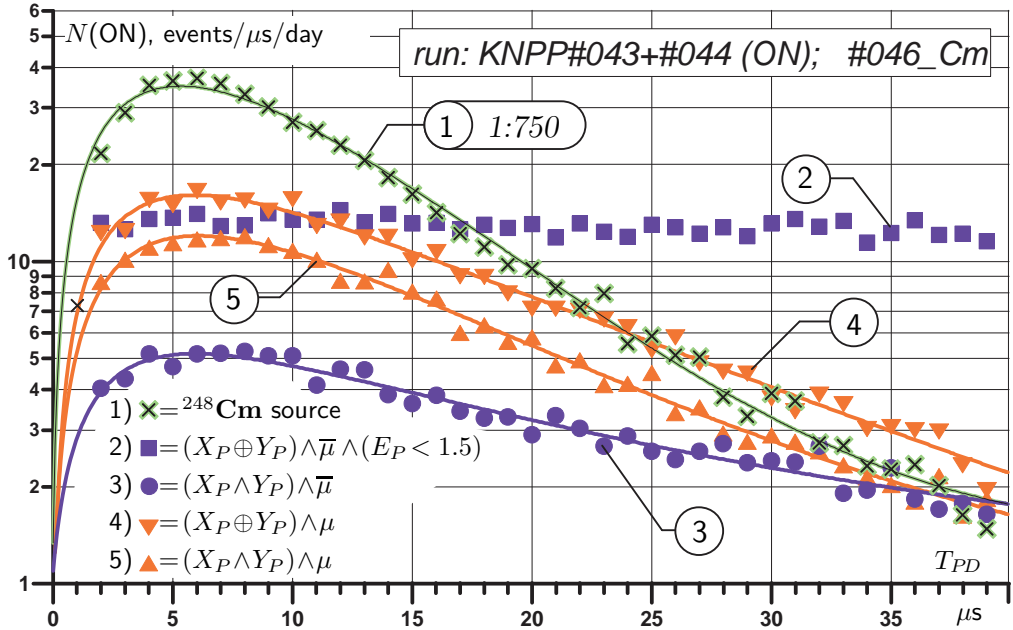}
 \caption{\footnotesize The $T_{PD}$ time distribution measured with the $^{248}$Cm source compared to the neutrino-like events detected under the operating reactor.}\label{Fig.Times(R44)}
 \end{figure}

The latter statement is confirmed by the flat $T_{PD}\,$ time distribution of the events with a low Prompt energy ($E_P\in[0.5-1.5$]~MeV) and ($X_P\!\oplus\! Y_P$) logic represented by curve 2 in Fig.~\ref{Fig.Times(R44)}. True IBD time curve 3 of the same figure corresponds to the lowest row of Table~\ref{Tab.Nu(ON-OFF)} and agrees with the expectation well. Its rate is significantly less sharp than that of curve 1 which was measured with the $^{248}$Cm source. Curves 4 and 5 correspond to the events associated with muons. The steepness of their rate is somewhat between that of curves 1 and 3, which reflects the intermediate neutron multiplicity ($k\simeq1.6$) for the muon-induced reactions in copper and lead shielding.

\begin{figure}[ht]
 \input{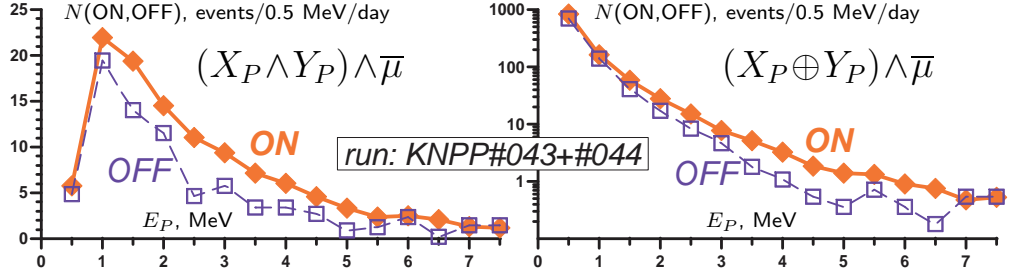}
 \caption{\footnotesize The $E_P$ energy spectra of the neutrino-like events detected at the reactor ON and OFF periods.} \label{Fig.Ep(ON+OFF)}
 \end{figure}

Energy spectra of the neutrino-like events detected at the reactor ON and OFF periods and corresponding to the alternative ($X_P\!\wedge\! Y_P$) and ($X_P\!\oplus\! Y_P$) logic are shown in Fig.~\ref{Fig.Ep(ON+OFF)}. The ON spectra demonstrate a non-zero positive excess over the OFF ones in both cases, but their shapes are very different for the reason mentioned above (the low-energy part of the ($X_P\!\oplus\! Y_P$) spectra is polluted with random $\gamma$-$n$ coincidences).

\begin{figure}[hb]
 \input{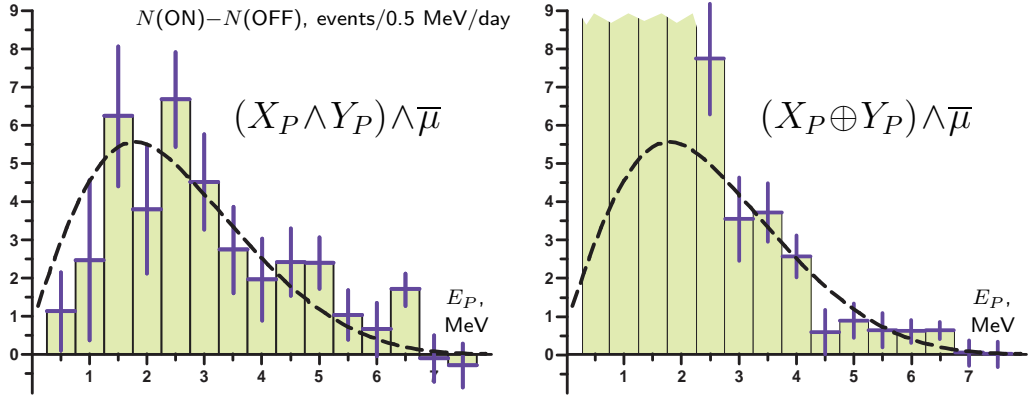}
 \caption{\footnotesize The differential $E_P$ energy spectra of the neutrino-like events detected by DANSSino. The dashed curve represents a typical example of the IBD positron energy spectrum simulated for the $^{235}$U fission \cite{Nu-spectra-old,Nu-spectra-new,Nu-evolution} and normalized to the number of observed events.}\label{Fig.Ep_Nu(ON-OFF)}
 \end{figure}

 Nevertheless, assuming that the OFF data correspond to a pure background, one can build an energy spectrum of IBD positrons as a difference $S$=$N$(ON)--$N$(OFF). As an example, $E_P$ spectra built under the ($X_P\!\wedge\! Y_P$) and ($X_P\!\oplus\! Y_P$)  requirements are shown in Fig.~\ref{Fig.Ep_Nu(ON-OFF)}.

 In spite of poor statistics and merely illustrative character of the spectra, the first of them and the higher-energy part of the second are in very good agreement with the GEANT4 MC simulations taking into account the measured detector parameters (photo-electron yield, attenuation of the light signal along the strip, etc.). For example, MC simulations predict 75 events per day for run \#044 while 70$\pm$5 events were detected. This agreement gives additional confirmation that we indeed observe IBD events.
\section{Summary}
As a result of numerous tests performed with the DANSSino detector, the following conclusions can be drawn:
\begin{itemize}
\item The most important background under the WWER-1000 reactor originates from fast neutrons produced by cosmic muons in high-Z surroundings. Therefore, one should not place heavy materials inside the neutron moderator.
\item Efficiency of the muon-veto system should not be less than 95-97\%. To reach this level, it is planned to install a double layer of scintillator plates with a lower threshold and use them in the coincidence mode.
\item In spite of the small size, big edge effects, incomplete passive and active shielding, and extremely simplified acquisition system, DANSSino is able to detect reactor antineutrinos with the signal-to-background ratio around unity and the efficiency at a level of 10\%, which is in good agreement with the MC simulations.  As the full-scale DANSS detector is of much higher volume, its response function is expected to be considerably better and efficiency significantly higher ($\simeq$70\%) because of a lower relative contribution from the edge detector parts (fewer neutrons and $\gamma$-rays after $n$-capture in Gd would leave the sensitive volume without detection, and the {\sl hard} thresholds could therefore be increased above 2--3~MeV without loss of efficiency). Together with the additional information from the individual photo sensors providing the space pattern of each event, it will allow to suppress the background down to a negligible value.
\item Operation of such detectors at a shallow depth with overburden less than 10-20 m w.e. seems to be questionable, as the neutron component of cosmic rays cannot be tagged by any veto system and produces a signature very similar to the IBD but outnumbers it by orders of magnitude.
\end{itemize}

\section{Acknowledgements}
The work was supported in part by the JINR (grant 13-202-05), Russian Foundation for Basic Research (grants 11-02-01251 and 11-02-12194), the Russian Ministry of Education and Science (grants 8174, 8411, 1366.2012.2) and corporation Rosatom (state contract  H.4x.44.90.13.1119).
The authors are grateful to the staff of the Kalinin Nuclear Power Plant and especially to the KNPP Directorate for providing a possibility of performing research measurements extremely close to the reactor core and to the KNPP Radiation Safety Department for the constant technical assistance.

\end{document}